\documentclass[draft]{agujournal2019}

\usepackage[utf8]{inputenc}
\usepackage{siunitx,xfp}
\usepackage{calc}
\usepackage{xassoccnt}
\usepackage{catchfile}
\usepackage{amsmath}
\usepackage{amssymb}
\usepackage{amsfonts}
\usepackage{subcaption}

\def\wordcountfile#1#2{
  \immediate\write18{detex '#1.tex' |wc -w > '\jobname.wc'}

  \newread\wcinput
  \openin\wcinput=\jobname.wc

  \begingroup
    \endlinechar=-1
    \read\wcinput to \localline

    \expandafter\xdef\csname #2\endcsname{\localline}
  \endgroup
  \closein\wcinput
}

\NewTotalDocumentCounter{totalfigures}
\NewTotalDocumentCounter{totaltables}
\DeclareAssociatedCounters{figure}{totalfigures}
\DeclareAssociatedCounters{table}{totaltables}

\usepackage{url} 
\usepackage{lineno}
\usepackage[inline]{trackchanges} 
\usepackage{soul}

\draftfalse

\journalname{Geophysical Research Letters}

\begin{document}
%
%


\title{Understanding Non-Gaussian Chorus Wave Statistics via the Benjamin-Feir Index}

%
%




\authors{D.J.Ratliff\affil{1}, O.Allanson \affil{2,3}, D. Rasinskaite \affil{1,4}, J. Stawarz \affil{1}, C. E. J. Watt \affil{1}, S. Chakraborty \affil{1}, A. W. Smith \affil{1} 
}


\affiliation{1}{Department of Mathematics, Physics and Electrical Engineering, Northumbria University, Newcastle upon Tyne, NE1 8ST, United Kingdom}
 \affiliation{2}{Space Environment and Radio Engineering, School of Engineering, University of Birmingham, Birmingham, B15 2TT, UK}
 \affiliation{3}{Faculty of Environment, Science and Economy, University of Exeter, Exeter, UK}
\affiliation{4}{Department of Physics, University of Helsinki, Helsinki, Finland}




\correspondingauthor{D.J. Ratliff}{daniel.ratliff@northumbria.ac.uk}



\begin{keypoints}
\item We derive a wave activity index for non-Gaussian frequency spectra.
\item Global maps suggest non-Gaussian wave activity regions consistent with observations.
\item Our wave action model produces frequency spectra in good agreement with Van Allen Probe measurements.
\end{keypoints}

%
%

%
%


\begin{abstract}
We derive an extended wave action model for equatorial chorus waves, identifying a wave activity index (a version of the Benjamin-Feir index, BFI) which indicates non-Gaussian frequency spectra emerge when BFI$>$0.5. Global maps of this index indicate the night and dawn sectors ($0<{\rm MLT}<9)$ of the magnetosphere as the primary region for non-Gaussian wave statistics. Comparisons with Van Allen Probe A burst mode data demonstrate qualitatively good fits, and despite being parameter-free our framework produces skill scores which indicate it is comparable with conventional least-squares Gaussian approaches whilst being able to produce asymmetric and multimodal chorus frequency distributions. Numerical evidence suggests the BFI$>0.5$ threshold persists in our wave action framework. This work ultimately establishes the first wave activity index that distinguishes Gaussian and non-Gaussian wave scenarios from first principles, providing the groundwork for a threshold-based quantification for use in space weather modelling.
\end{abstract}

\section*{Plain Language Summary}
The chance of encountering intense, potentially dangerous, electromagnetic wave activity in near-Earth space depends on many factors that arise from the local plasma environment. How these factors might contribute to these statistics is currently not fully understood. We use modelling approaches first developed in ocean wave forecasting to explore which factors may contribute to extreme wave statistics. This reveals an index whose value suggests when a collection of waves may remain predictable (and have low probability of large amplification) or when they may become increasingly and abnormally intense. Exploring the value of this index using existing datasets, we find the most abnormal wave activity happens for intense waves on the nightside of the Earth. We use computer simulations to  compare our model’s predictions to those directly taken from spacecraft and find good agreement in the predicted shape of the frequency spectrum and reasonable agreement with its statistical measures. We additionally find that the new terms in the model account for the skewed and non-symmetric frequency spectra that allow our predictions to be as accurate as they are.


%
%

%


%
%
%
%

\section{Introduction}

Chorus wave activity in the Earth's magnetosphere, particularly its impact on the particle populations in near-Earth space, remains an important facet of modern space weather \cite{Thorne-2010,Thorne-2013}. These waves can accelerate electrons via gyroresonant wave-particle interactions (WPIs) to higher energies~\cite{Lejosne-2022,Summers-2002}, potentially reaching dangerous levels (0.5 MeV to multi-MeV) that pose risks to space infrastructure such as satellites. The ability of chorus to do this depends on several factors, including the local plasma environment and wave properties, which makes predicting chorus-driven space hazards on the shorter term an ongoing challenge. 

Significant effort has gone into deriving statistical models which capture typical chorus properties that can be incorporated into operational models. Most notably, the Gaussian approximation has emerged to describe the power spectral density of chorus activity and has become a typical feature in radiation belt modelling~\cite{Glauert-2005,Horne-2013,Horne-2018}. The success and tractability of this paradigm has motivated several chorus wave models based on multi-year and/or multi-mission data to supply wave parameters based on geomagnetic and environmental conditions \cite{Agapitov-2015,Agapitov-2018}. However, the observed wave frequency spectrum can be significantly non-Gaussian - being skewed, heavy-tailed and/or multimodal \cite{bunch-2013,wang-2019,li-2015} and investigations at the diffusive scale have demonstrated that such spectra can
lead to energy and pitch angle diffusion coefficients orders of magnitude different than their Gaussian
counterpart \cite{li-2015,ni-2019,wang-2019,wang-2024,wong-2024}. So whilst it is becoming clear what the impacts of non-Gaussian wave statistics are, their origins and drivers remain unclear and largely absent from the literature. 

The origin of non-Gaussian wave statistics can be understood via another geophysical context which considered non-Gaussian problems before - oceanography. Motivated by extreme wave statistics within maritime safety, several authors worked to produce frameworks to explore the dynamics and statistics of ocean wave spectra, resulting in operational wave action models~\cite{Tolman-2009,ECMWF-2019,Janssen-2003,Onorato-2001}. Such models identify a nondimensional parameter associated with wave-wave energy exchanges - the Benjamin-Feir index (BFI) - which indicates when wave statistics are expected to be Gaussian (and follow near-linear wave statistics) or become increasingly non-Gaussian and support extreme wave activity, and thus remains a key indicator of dangerous wave activity and a crucial operational output in oceanic forecasting.
Similar wave action approaches have emerged independently within works on intense whistler-wave spectra, where wave-wave energy exchange is referred to as nonlinear scattering in the context of plasmas and radiation belt physics~\cite{Crabtree-2012,Ganguli-2010,ganguli-2012,ganguli-2019,soto-2023,tejero-2015,tejero-2016}. However such work has not focussed on the non-Gaussian nature of the power spectra such models can produce, nor the regimes in which it can be produced. This presents a natural opportunity to use these frameworks for chorus to understand the onset of non-Gaussian statistics within chorus waves.

This paper aims to identify an analogous index that may differentiate between when and where Gaussian and non-Gaussian chorus wave statistics should be expected in the magnetosphere. We achieve this by considering a wave action framework that can be simplified by considering equatorial parallel-propagating, narrow-banded chorus waves, which admits a version of the BFI for chorus, allowing us to explore the magnetospheric variability of this index across a range of $AE$ activity indices. We find that the largest values of the BFI index happen within the night and dawn sectors ($0<{\rm MLT}<9$) during active times, correlating with existing observations of intense chorus activity.

This study also finds that we must consider additional terms in our wave action model to capture chorus phenomenology, specifically those shown to generate tonal behaviour \cite{Ratliff-2023}, leading to a wave action model we refer to as the Whistler Action Model (WhAM), derived in section \ref{sec:WhAM}. The simulations of the WhAM undertaken in section \ref{sec:VAPs} demonstrate that the inclusion of such terms causes additional asymmetries in the power spectrum that become more pronounced with increasing BFI. These extensions are vindicated by comparisons between WhAM predictions and observed wave spectra from the Van Allen probes, finding good qualitative agreement, low skill scores when compared to existing Gaussian paradigms and events that highlight the need for energy sources and sinks within the WhAM framework.

\section{Wave Action Modelling of Chorus Waves}
\label{sec:WhAM}
This paper considers equatorial parallel propagating (field aligned) chorus in a uniform magnetic field of strength $B_0$ under a wave action approach. Such models balance linear wave propagation with the energy exchange between waves through nonlinear processes to explore how the wave's power spectrum evolves dynamically, though the nonlinear component of the dynamics is often difficult to determine and analyse without some initial simplifying assumptions. For whistler-mode waves like chorus, a weak turbulence framework is appropriate, meaning the nonlinear interactions can be expressed as products of the wave amplitudes across the spectrum.  Whilst the leading order nonlinearity would be expected to be quadratic in general~\cite{Galtier-2000,David-2022}, previous work has shown wave-wave energy exchanges for parallel propagating chorus are cubic and are mediated by non-resonant ponderomotive WPIs~\cite{Ratliff-2023,Krafft-2018,Karpman-1989,Karpman-1990,Ganguli-2010,Crabtree-2012}. Under these assumptions and observations, the principal wave action model is
\begin{equation}\label{eqn:weak-turb}
i\frac{d B_w(k,t)}{dt}-\omega(k) B_w = \sum_{k_1+k_2-k_3-k = 0}T(k_1,k_2,k_3,k)B_w(k_1,t)B_w(k_2,t)B_w^*(k_3,t)\,,
\end{equation}
where $k$ is the field aligned wavenumber, $\omega(k)$ is the wave frequency arising from the whistler-mode cold plasma dispersion relation~\cite{Stix-1992}, the asterisk denotes complex conjugation.

The function $T$ in (\ref{eqn:weak-turb}) mediating the nonlinear terms represents the wave-wave energy transfer function~\cite{Newell-2011}, whose form is typically involved and does not lend itself readily to analysis. To make progress we utilise the fact that chorus are narrow-banded~\cite{Santolik-2003,Santolik-2008} to restrict our focus close to the peak of the wave activity, centred at wavenumber $k_0$ (or equivalently, frequency $\omega (k_0)\equiv \omega _0$). This motivates introducing the following scalings:
\[
\begin{split}
k = k_0+\sigma_k p\,, \quad B_w =\sqrt{ \langle |B_w|^2\rangle}& \,B\,, \quad \tau = \frac{1}{2}\omega''(k_0)\sigma_k^2\,t\,, \quad {\rm where} \quad  \sqrt{ \langle |B_w|^2\rangle} \,, \ \sigma_k \ll 1\,,
\end{split}
\]
and primes denote differentiation with respect to $k$.
The statistical parameters $\sigma_k$ and $\sqrt{ \langle |B_w|^2\rangle}$ characterise the spectral width and root-mean square intensity of the wave spectrum respectively.
This facilitates the Taylor expansion of both $\omega$ and $T$. Typically only terms up to $p^2$ of $\omega$ and the leading order value of $T$ are retained, leading to a model that retains a $p\to -p$ symmetry, forbidding skewed power distributions and movements in the spectral mean. This is inappropriate for chorus whose frequency spectra can be asymmetric, and so we continue each series expansion one order higher, verifying the neglected terms are indeed smaller than what is retained for lower band chorus. After removing the background and group velocity terms (equivalent to changing to a frame of reference moving with the group velocity) we obtain the simplified dynamical equation:
\begin{equation}\label{eqn:WhAM}
    i \frac{d B(p,\tau)}{d\tau}-(p^2+D_3p^3)B+{\rm BFI}^2\sum \delta_{p_1+p_2-p_3-k}(1+\mathcal{S} p_1) B(p_1)B(p_2)B^*(p_3) = 0\,.
\end{equation}
where $\delta$ is the Kronecker delta,
\[
\begin{split}
{\rm BFI} &= \sqrt{2} \, \frac{\sqrt{\langle|B_w|^2\rangle}}{B_0}\frac{k_0}{\sigma_k} \sqrt{-\,\frac{T(k_0,k_0,k_0,k_0) B_0^2}{\omega''(k_0)k_0^2}}\,,\\[3mm]
T(k_0,k_0,k_0,k_0) &= \frac{\Omega}{B_0^2}\frac{\omega_{pe}^2\Omega (\Omega-\omega_0)^2\big(\Omega(2\omega_0^2+\omega_{pe}^2)-2\omega_0^3\big)}{(\Omega\omega_0 - \omega_0^2 + \omega_{pe}^2)\big(\omega_0(\Omega-\omega_0)^2 + \frac{1}{2}\omega_{pe}^2\Omega\big)^2}\\
& \hspace{3cm}\times\left(\frac{c^2}{v_g(k_0)^2-v_{th,c}^2}+\frac{c^2}{v_g(k_0)^2-v_{th,s}^2}\right)\,, \\[3mm]
D_3 &= \frac{\sigma_k}{3}\frac{\omega'''(k_0)}{\omega''(k_0)}\,,\\[3mm]
\mathcal{S} &= \frac{\sigma_k}{k_0} \frac{(2\omega_0-\Omega)(2\omega_0(\omega_0-\Omega)-\omega_{pe}^2)}{2\omega_0(\omega_0-\Omega)^2+\omega_{pe}^2}\,,
\end{split}
\]
and $\Omega,\, \omega_{pe},\,B_0,\,v_g(k_0),\,v_{th,c},\,v_{th,s}$ are the (unsigned and non-relativistic) electron gyrofrequency, electron plasma frequency, background magnetic field strength, group velocity and thermal velocity of the cold and seed (those of energies in the eV and keV range respectively) electron species respectively. It is this model, which we refer to as the Whistler Action Model (WhAM) and will be used in this paper to explore the wave distributions of chorus.

This is now a problem of three parameters. The first, (BFI),  characterises the nonlinear wave-wave interactions  and is referred to as the Benjamin-Feir index. It is the most well-documented of the parameters in this model and has been demonstrated that extreme, non-Gaussian wave activity emerges whenever this index exceeds $1/2$ \cite{Janssen-2003,Janssen-2007,Onorato-2001,Serio-2005}. We note that although the theoretical transition was shown to be $BFI=1$, numerical studies show that non-Gaussian statistics emerge prior to this point~\cite{Janssen-2003}. Conversely, index values below this threshold indicate Gaussian wave spectra are stable, and BFI\,$\in \mathbb{C}$ (equivalently, $T(k_0,k_0,k_0,k_0) \omega''(k_0)>0$) results in a defocussing wave dynamic where wave energy disperses rather than collects in discrete, large amplitude wave packets. As such, it has since formed a key output in modern ocean wave action models as an indicator of dangerous wave activity \cite{Tolman-2009,ECMWF-2019}, and therefore will be the primary parameter we explore within this paper to assess its effect on the statistics of chorus waves.

The remaining parameters $D_3$ and $\mathcal{S}$ are referred to as Dysthe (or modified) correction terms~\cite{dysthe-1979,trulsen-1996}. The former represents higher order further (linear) dispersive effects into the wave model. The latter reintroduces more of the full wave-wave energy exchange into the wave action framework, and this specific term has been shown to account for frequency shifting behaviour in chorus waves and is able to account for the rising tone structures synonymous with intense chorus waves~\cite{Ratliff-2023}. Therefore, these model additions are expected to modify the resulting shape and odd-order statistical moments (e.g. the mean and the skewness) but not strongly affect the threshold for the emergence of non-Gaussian statistics. In light of this we do not perform a parameter scan of these coefficients as part of this paper, and simply utilise the values at the corresponding plasma state to the desired BFI value.

\subsection{The Global Picture of the Benjamin-Feir Index}

\begin{figure}[!ht]
    \centering
  \includegraphics[width=\textwidth]{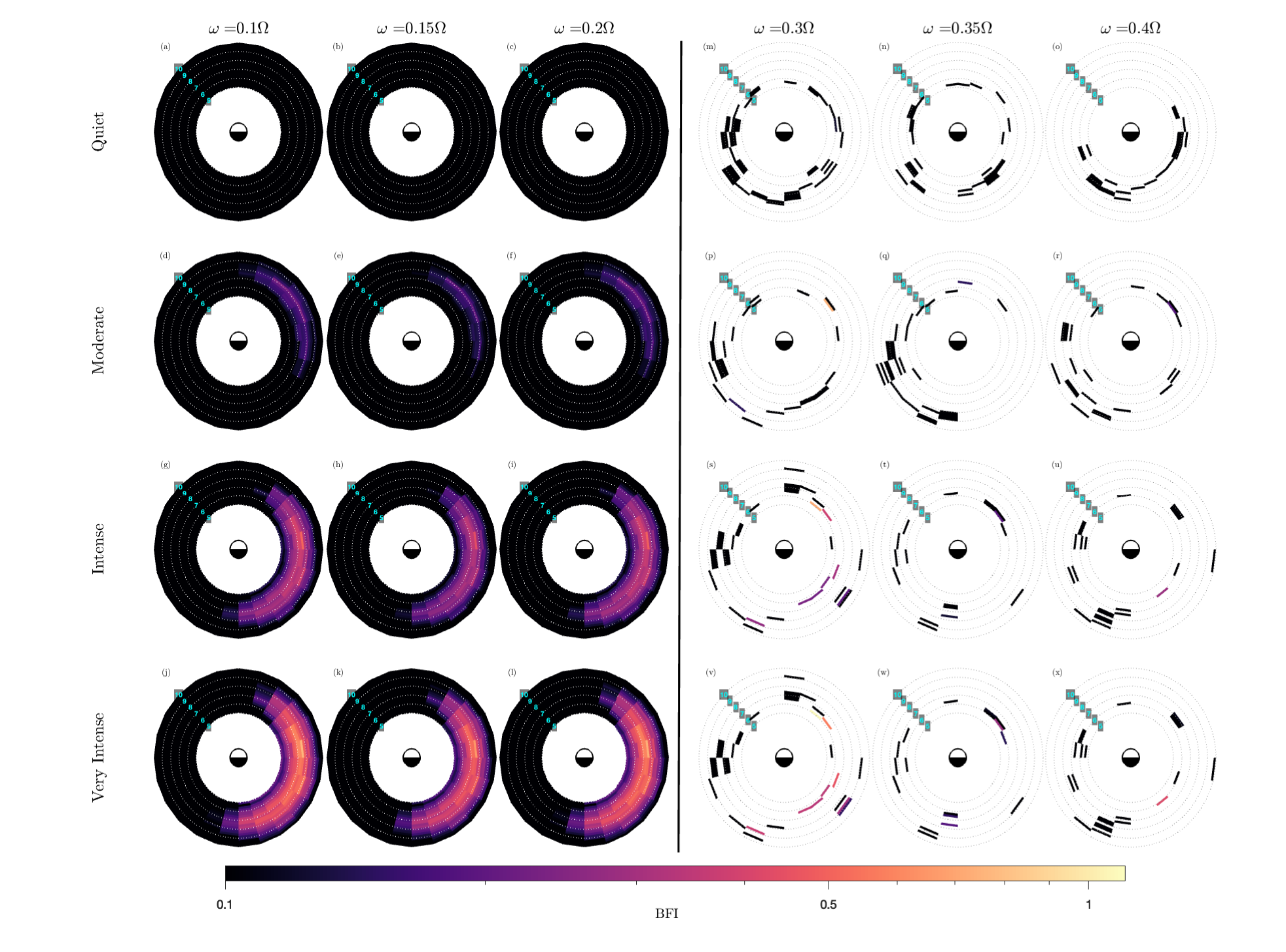}
    \caption{Global maps of the BFI index in $5<L<10$ across four chorus activity ranges (10pT$<|B_W|<$30pT (quiet),  30$<|B_W|<$100 (moderate), 100$<|B_W|<$300 (intense) and $|B_W|>$300 (very intense)) and across peak frequencies. Maps for $\omega<0.25\Omega$ ((a)-(l)) are computed for electrons with warm temperature 30 keV and 5 keV for $\omega>0.25\Omega$ ((m)-(x)), with these regimes separated by the vertical line.}
    \label{fig:fig1-BFI-Map_lower}
\end{figure}


The established relationship between the BFI and non-Gaussian wave statistics in the field of oceanography motivates a study of this index globally to infer where the model (\ref{eqn:WhAM}) suggests non-Gaussian wave statistics should be expected. We do so by constructing  global distributions of the BFI in $5<L<10$ utilising statistics and models synthesised from the THEMIS~\cite{Angelopoulos-2008} and Van Allen Probes~\cite{Mauk-2013} missions, which we outline below.

Our maps will consider 4 levels of chorus wave activity outlined in \cite{Li-2011}: $10\ {\rm pT}<|B_W|<30 \ {\rm pT}$ (quiet),  $30\, {\rm pT}<|B_W|<100 \, {\rm pT}$ (moderate), $100\, {\rm pT}<|B_W|<300 \, {\rm pT}$ (intense) and $|B_W|>300 \, {\rm pT}$ (very intense). The wave intensity used for the BFI calculation is taken to be the midpoint of each bin, and we take an intensity of $300$pT for the final bin. 
We globally distribute the magnetic wave intensity in each case according to the function
\begin{linenomath*}
\[
\langle |B_w|^2\rangle(L,\varphi) = I_{max}+(I_{max}-100)\exp\left(-\frac{(L-L_{peak})^2}{8}\right)\exp\left(\cos(\varphi-\varphi_{peak})-1\right)\,,
\]
\end{linenomath*}
choosing the maximum intensity $I_{max}$ as the aforementioned midpoint bin values and the position of the peak is determined by the pair $(L_{peak},\varphi_{peak})$. For quiet and moderate wave activities, $(L_{peak},\varphi_{peak}) = (7,8)$ whilst for higher intensities $(L_{peak},\varphi_{peak}) = (6.5,4)$ \cite{Li-2009,Meredith-2020}. The final aspect of the wave spectrum's statistical properties, the spectral width parameter, is chosen to be $16\%$ of the carrier frequency, in line with the literature \cite{Li-2008,Burtis-1976,Agapitov-2018}, which we convert to a wavenumber width via the differential relationship $\sigma_\omega = c_g\sigma_k$ \cite{Chabchoub-2016}.

For the plasma environment, the number densities and values of the ratio $\omega_{pe}/\Omega$ are taken from models determined from THEMIS \cite{Li-2009}, using each level of $AE$ activity for the corresponding wave intensity classification (using the high $AE$ index values for both intense and very intense cases). We can determine a magnetic field strength $B_0$ from these. For the electron temperatures, we prescribe 5eV for the cold population and a seed population temperature of $10$ keV \cite{Rasinskaite-2025}. Whilst other studies show that this can vary up to 50 keV for seed populations \cite{Denton-2010} this does not qualitatively impact our BFI global picture, and only very weakly impacts it quantitatively.
We present our maps over a range of frequencies in the lower chorus band to ascertain the frequency dependence of this index. We note that the frequency $\omega = 0.25\Omega$ is omitted from our parameter scan as for dense plasmas $\omega_{pe}/\Omega>1$ there is a singularity in the BFI due to $\omega''(k_0)$ vanishing at precisely this frequency. We also do not visualise points where BFI $\in \mathbb{C}$.

The resulting maps of the BFI for various activity indices are presented in Figure \ref{fig:fig1-BFI-Map_lower}. For the lower frequencies ($0.1\Omega\leq \omega \leq 0.2 \Omega)$ we observe that quiet chorus activity leads to very low BFI values, suggesting the frequency spectra in such cases are expected to be Gaussian. Conversely, for intense and very intense chorus order 1 BFI values are observed in the night/dawn sector ($0<{\rm MLT}<9$). Only quantitative differences exist between the intense and very intense categories due to only differing by wave intensity. Interestingly, the highest BFI values in these cases do not coincide with the prescribed peak of the wave statistics, demonstrating that the wave intensity statistics are not the principal factor in the size of the BFI indices here. 
For the higher peak frequencies in Figure \ref{fig:fig1-BFI-Map_lower} ($0.3 \Omega\leq \omega \leq 0.4\Omega$) we find a much more inconsistent BFI map due to few grid points admitting a real-valued BFI. Of those which do, these either occur in the dusk sector($12<{\rm MLT}<18$) with a very low BFI value, or in the dawn sector with values expected to correspond to non-Gaussian wave statistics.

\section{Case Study: Van Allen Burst Mode Data}
\label{sec:VAPs}

\begin{figure}
    \centering
    \includegraphics[width=\textwidth]{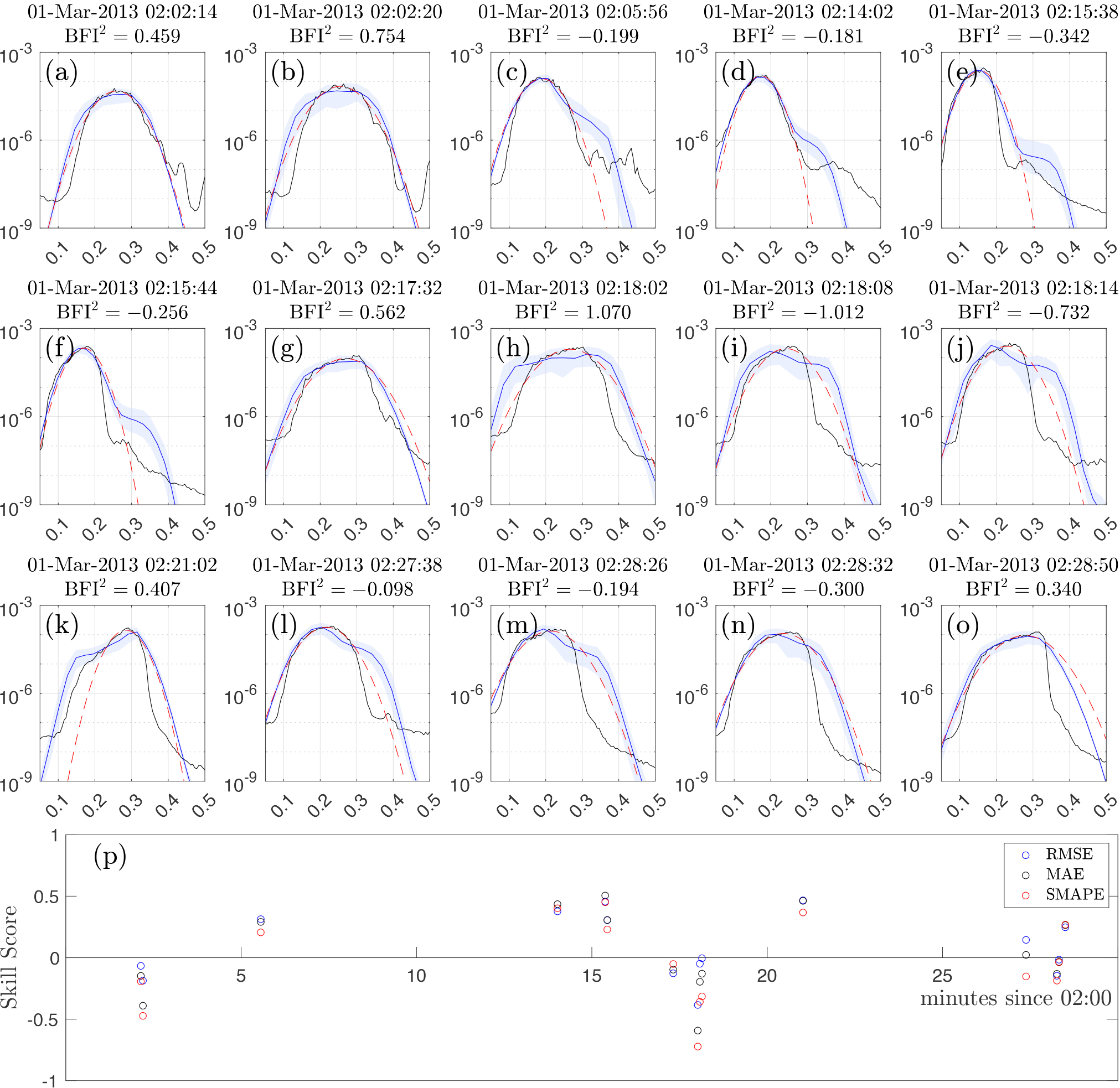}
    \caption{Panels (a)-(o): Comparisons between the frequency spectrum measured by the Van Allen probes (black solid), the corresponding WhAM simulations (blue solid) and a least squares Gaussian fit (red dashed) across several burst-mode intervals from $1^{st}$ March 2013 02:00 UT. The timestamp and the square of the BFI indices for each measurement appears in each title. The blue shaded region denotes the 95\% confidence intervals for the Monte Carlo simulations. Panel (p): Skill scores for the WhAM spectrum versus the least squares Gaussian fit plotted against time of measurement.}
    \label{fig:VanAllenComp}
\end{figure}

We now use this framework and insight to determine how the WhAM framework performs against in-situ wave spectra.
We utilise the Van Allen Probe continuous burst mode dataset obtained from the Electric and Magnetic Field Instrument Suite and Integrated Science (EMFISIS) instrument \cite{Kletzing-2013} to generate our wave spectra for 6 second intervals. We focus our discussion on the observations from Probe A taken at 02:00 UT on $1^{\rm st}$ March 2013, owing to periods where the wave power is mostly conserved (in contrast to intermittent power due to rising tones) which better emulate the simulations performed on (\ref{eqn:WhAM}), which assumes conserved wave power. We select intervals where the root mean square error between the wave power measurements $P_i$ and the mean value of the power on the interval $\overline P$ is less than the average wave power:
\[
\frac{\sqrt{\frac{1}{N}\sum_{i=0}^6(P_i-\overline P)^2\, dt'}}{\overline P}<1\,.
\]
We find 15 of the 30 burst mode datasets satisfy these requirements  and are timestamped in Figure \ref{fig:VanAllenComp}. These datasets allow us to construct the frequency spectrum and the associated BFI for each six second interval. The selected intervals admit both real and imaginary BFI values, allowing us to assess this index and the WhAM framework in both the focussing and defocussing regimes.

Simulations of the WhAM (\ref{eqn:WhAM}) are performed using an ensemble modelling approach following \citeA{Janssen-2003} with minor alterations. For each burst mode event, we initialise a Gaussian wave spectrum across a wavenumber domain which spans the chorus lower band ($0.1\leq \omega/\Omega <0.5$) with random phases and simulate (\ref{eqn:WhAM}) for 60 nondimensional wave periods (corresponding to physical/dimensional times between 6s and 18s), timestepping using an exponential time differencing scheme with Runge-Kutta timestepping method \cite{Cox-2002,Kassam-2005} with spatially periodic boundary conditions. The average power and intensity distributions are obtained for each ensemble run, excluding the first 10\% of the run (to allow for phase mixing to occur) which are then converted to frequency space via the differential relationship $\delta \omega  = c_g \delta k$ ~\cite{Chabchoub-2016} and averaged to obtain an overall frequency distribution for the given plasma and wave parameters. Supplementing the spectrum's statistical parameters we obtain from the burst mode distribution is the associated EMFISIS level 4 density data, used to obtain the relevant plasma parameters to construct the WhAM model \cite{Kurth-2015}. The only unavailable parameters are the temperatures of the electron species, and so we make the assumption that the cold species has a temperature of 5 eV and the warm population has one of $30\,$keV \cite{Denton-2010,Rasinskaite-2025}. The resulting BFI values are not sensitive to these choices for the temperatures as the resulting thermal velocities are not close to $v_g(k_0)$.

The result of our simulations and the comparison to burst mode events and a least-square Gaussian fit of the observation, consistent with the conventional approach of the field \cite{Glauert-2005,Horne-2013}, are visualised in Figure \ref{fig:VanAllenComp}. We first observe that we are unable to test our BFI threshold from this data due to a low number of BFI$<1/2$ measurements, and so we must test this threshold synthetically (see \S \ref{sec:BFIScan}).
The majority of cases give good agreement to the observed distributions, and perform comparably to the least-squares fit (which is designed to minimise the root-mean square error). 
We formalise the WhAM's ability to predict the spectral shape compared to a Gaussian fit using skill scores~\cite{Murphy-1988}, presented here are three metrics on the logarithm of the distributions - the root-mean square (RMSE), mean absolute error (MAE) and symmetric mean absolute percentage error (SMAPE)~\cite{Morley-2018,Armstrong-2010,Liemohn-2021}. We find that WhAM provides a better model in many cases, with some clear instances of worse performance. One such interval is that spanning subplots (h)-(j), however the timestamps suggest that rapid temporal variation may be responsible for poor model agreement. This is supported by the large change in BFI (from ~1 to -1) due to a significant changes in the wave's mean frequency over a short timescale, accompanied by variations in the root-mean square wave power which are also present in the event spanning panels (a) and (b). The latter suggests missing phenomenology within the WhAM framework necessary to capture the process occurring in these windows. Given the timescale over which this occurs, we speculate that this is due to energy entering the system due to resonant WPIs, and the WhAM model does not possess energy sources or sink terms. 
The inclusion of energy sinks would act to suppress the heavier-than-observed tails seen in the WhAM outputs where ${\rm BFI}^2<0$ and likely improve model predictions.

Regardless, the model still seems to accurately capture the transfer of wave energy across the spectrum in all cases, even those with poor skill scores. Further, it should be noted that within this case study, all cases with a negative ${\rm BFI}^2$ value have dominant peaks well described by Gaussian fits, in keeping with BFI threshold paradigms from oceanography. However the observational data suggests even chorus waves in defocussing cases can also be weakly non-Gaussian, in the case of this event due to small multimodal peaks emerging. Overall, these results, observations and skill scores are  indicative that the wave-wave energy exchange terms due to nonresonant WPIs are an important part of the physics which are shown here to produce non-Gaussian chorus wave spectra. Therefore, such WPIs should be considered more closely in the study of chorus wave spectra going forward.

\subsection{The Influence of the BFI on Spectral Structure}
\label{sec:BFIScan}
Our results demonstrate that a variety of non-Gaussian spectral shapes may be produced by the WhAM, with similar structures of varying prominence emerging across a range of BFI indices, best exemplified by Figure \ref{fig:VanAllenComp} panels (c) and (e). This motivates a more controlled investigation of the role of the BFI in spectral shape. For each of these cases, we retain the plasma and wave parameters used to generate each panel and then control the RMS intensity so that $|{\rm BFI}| = 0.25, 0.5,\,1,\, 1.25$. This keeps the linear term $D_3$ fixed across the scan in BFI value. We simulate (\ref{eqn:WhAM}) as described previously for these BFI values and produce the resulting spectra in each case. 

\begin{figure}[!ht]
    \centering
\includegraphics[width=\textwidth]{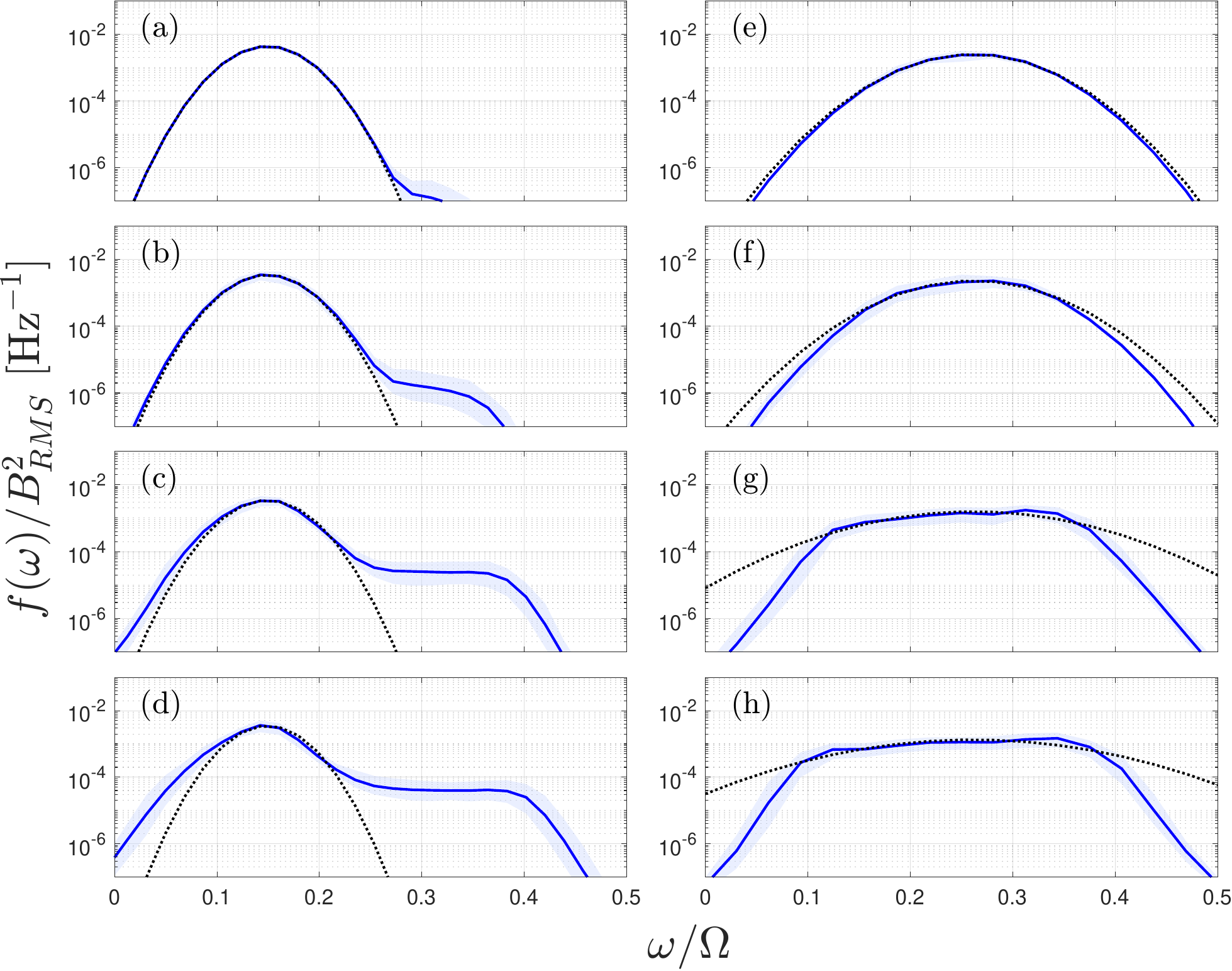}
    \caption{Scans across the BFI index for cases (c) (left column, BFI imaginary) and (e) (right column, BFI real) from \ref{fig:VanAllenComp}. Rows correspond to $|{\rm BFI}| = 0.25,\,0.5,\,1,\,1.25$. Blue lines present the results of simulating (\ref{eqn:WhAM}) with light blue shaded regions denoting the 95\% confidence intervals and black dotted line denoting the least-squares Gaussian fit to the Monte Carlo distribution.}
    \label{fig:BFI_scans}
\end{figure}

Our results in Figure \ref{fig:BFI_scans} make it clear that the BFI=$1/2$ remains an approximate threshold for non-Gaussian frequency spectra (illustrated by the least squares Gaussian fit closely modelling simulated distribution below this value). As seen in the case study of this paper, there is emergent non-Gaussian behaviour for the BFI$\in \mathbb{C}$ in a departure from the oceanographic case, where a second peak in the distribution emerges below this threshold. Despite this, the Gaussian least squares fit remains a good fit for the dominant frequency peak. 

In both cases, the emergent shape of the non-Gaussian spectrum is accentuated for increasing BFI values, demonstrating the BFI index (and by virtue, the nonlinear component of (\ref{eqn:WhAM})) is a driver of the non-Gaussian shape of the frequency distribution, rather than the extended linear term of our WhAM framework. This points to the physical mechanism for emergent non-Gaussian chorus spectrum in this description being the non-resonant (ponderomotive) WPIs. This is an important observation -   gyroresonant WPIs are regularly attributed to the formation of non-Gaussian statistics, but such processes are entirely absent in our model. Our study here therefore motivates a re-assessment of non-resonant WPIs in chorus wave dynamics and their impact on chorus frequency spectra.

\section{Discussion}
\label{sec:conclusions}
This paper presents the WhAM framework, the first dynamical model we are aware of that can capture the origin and form of non-Gaussian statistics in chorus waves. The onset of non-Gaussian statistics is captured by the BFI, whose value delineates between Gaussian (BFI $\le 0.5$) and non-Gaussian (BFI $\geq 0.5$). In contrast with the existing narrowbanded literature, we find the additional features in the WhAM lead to skewed wave power distributions, and have also been shown to produce even multimodal distributions. Furthermore,  preliminary evidence from in situ data supports the ability of this framework, even in the narrowbanded approximation, to reproduce measured chorus spectra.

The potential utility of models like the WhAM in space weather forecasting lies within their use in tandem with energy and pitch angle diffusion coefficients. The chorus wave spectrum forms a key input into these coefficients, and having a framework which can generate wave statistics based on a given plasma environment from first principles will permit exploration of non-Gaussian statistics in these coefficients from a theoretical perspective for the first time. Moreover, our work is indicative that Gaussian wave statistics persist precisely when quasilinear theory (QLT) is operational. This connection can best be seen by defining the index
\[
{\rm KI} = \frac{\omega}{\sigma_\omega} \sqrt{\left(\frac{B_{RMS}}{B_0}\right)\left(\frac{v_\perp k_0}{\Omega}\right)} 
\]
where ${\rm KI}\ll 1$ is the classic Karpman condition for quasilinear theory~\cite{Karpman-1974,Tong-2019,Allanson-2024}. It is then apparent from the definition of the BFI that ${\rm KI}^2 \propto(\sigma_\omega /\omega)\, {\rm BFI}$ for a given plasma environment, and thus connects the onset of non-Gaussian statistics identified in this work with the breakdown of QLT. The violation in this context is due to non-resonant ponderomotive WPIs which are often neglected from QLT~\cite{Dodin-2022,Dodin-2024} but may have similar impacts as other non-resonant WPIs that also violate QLT \cite{Camporeale-2015,Allanson-2022,Yu-2025}.
Dedicated work towards how diffusion coefficients vary as a function of BFI, especially as quasilinear theory begins to break down, is therefore important to shed light on the role of ponderomotive WPIs in large scale particle dynamics. 


Our results demonstrate that several extensions to the WhAM should be made in the future to account for important phenomenology within the evolution of chorus. One key aspect is the energisation/dissipation due to gyroresonant interactions \cite{Omura-2008,Omura-2021}, which principally would form a forcing term within this model driven by particle distributions but may also present a coupled dynamical system, as the wave spectrum itself will drive the dynamics of the resonant particles. This interplay is perhaps the single most important next step for such models if they are to be developed for use in space weather applications. Further considerations (including Landau damping and field line curvature) can be incorporated by adapting existing work on wave action models for whistlers \cite{Ganguli-2010,Crabtree-2012}, using the work presented here as a foundation to explore the impact of these features on chorus statistics.

%
%

\section*{Open Research Section}
The numerical implementation of the WhAM in MATLAB and the data generated for the analysis in sections 2 and 3 can be found at [TBA]

\acknowledgments
DJR acknowledges the support of UKRI grant MR/Z506448/1. OA acknowledges support from the University of Birmingham and NERC Independent Research Fellowships NE/V013963/1 and NE/V013963/2. JS is supported by  the Royal Society University Research Fellowship URF/R1/201286. SC is supported by STFC Grant ST/V006320/1 and NERC Grants NE/V002554/2 and NE/P017185/2. AWS was supported by NERC Independent Research Grant NE/W009129/1.

%
%

\bibliography{references.bib}

%
%
%
%
%

\end{document}